\documentstyle[aps,preprint]{revtex}
\tightenlines
\begin{document}

\title{Cooperative dynamics in auditory brain response}

\author{J. Kwapie\'n$^1$, S. Dro\.zd\.z$^{1,2}$, L.C. Liu$^3$ 
                           and A.A. Ioannides$^{3,4,5}$}

\address{
$^{1}$ Institute of Nuclear Physics, PL--31-342 Krak\'ow, Poland,\\
$^{2}$ Institut f\"ur Kernphysik, Forschungszentrum J\"ulich,
D--52425 J\"ulich, Germany,\\
$^{3}$ Institut f\"ur Medizin, Forschungszentrum J\"ulich,
D--52425 J\"ulich, Germany,\\
$^{4}$ Physics Department,  The Open University, Milton Keynes, MK7 6AA, UK,\\
$^{5}$ Laboratory for Human Brain Dynamics,
BSI, The RIKEN Institute, Hirosawa 2-1, Wako-shi 351-01, Japan}
\date{\today}
\maketitle

\begin{abstract}

Simultaneous estimates of the activity in the left and right auditory cortex 
of five normal human subjects were extracted from Multichannel 
Magnetoencephalography recordings.  Left, right and binaural stimulation were
used, in separate runs, for each subject.  The resulting time-series of
left and right auditory cortex activity were analysed using the concept of 
mutual information.  The analysis constitutes an objective method to address
the nature of inter-hemispheric correlations in response to auditory 
stimulations.  The results provide a clear evidence for the occurrence of 
such correlations mediated by a direct information transport,
with clear laterality effects:  as a rule, the 
contralateral hemisphere leads by 10-20ms, as can be seen in the average 
signal.  The strength of the inter-hemispheric coupling, which cannot be 
extracted from the average data, is found to be highly variable from
subject to subject,  but remarkably stable for each subject. 

\end{abstract}


%\bigskip

\newpage

\section{Introduction}

Two emergent properties of complex systems are collectivity and chaos. Both 
properties are relevant for biological systems which some believe are 
balanced at the interface of collectivity and chaos~\cite{Kauf}. The brain 
itself has been described in these terms, particularly its tendency to
diversity and its ability of generating coherent patterns of activity, 
switching continuously from one to another. These properties are also 
expected to be very useful for describing how local cortical specialization 
is  efficiently co-ordinated by functional global integration 
mechanisms~\cite{Fris}.  Implicit in any such explanatory description is the 
brain's activity on various space and time scales. 

A quantitative understanding of the hierarchy of the underlying 
structures, both in space and in time, is of fundamental importance 
for a proper design of a unified theoretical model (for some attempts in this 
direction see for instance ref.~\cite{Somp}) relating local neuronal dynamics 
and global attributes of sensory processing. This, however, is an extremely
difficult problem since the conscious human brain is never at rest; 
central control of body function and regulation,  
fleeting thoughts and feelings, 
ensure that even in the most relaxed state a tapestry of regional activations 
is woven every instant.  Even the simplest of acts engages a multitude of 
areas in a way that varies even as the same task is repeated many times.
We have studied one of the simplest possible brain responses: the activity 
in the human auditory cortex, elicited by the presentation of simple tones,  
delivered regularly to one or both ears. Even in this very simple and 
artificial scenario animal~\cite{Aitk} and 
human studies~\cite{Goff,Laut} have shown that many 
different areas are involved.  Nevertheless, for this case the two auditory
areas are known to be active and prominent.  Magnetoencephalography 
(MEG)~\cite{Hama} is particularly appropriate in the present context, because 
activity from the auditory cortex is readily identifiable from both the 
average MEG signal~\cite{Hari,Sing} and in single trials~\cite{Liu1,Liu2}

Furthermore, since the two auditory 
cortices  are well separated on either side of the head,  the instrument at 
our disposal,  with two separate probes each with 37 channels, was ideal 
for mapping the magnetic signal: while one probe is sensing the signal 
over the left auditory cortex the other is sensing the signal over the 
right auditory cortex. With optimal sensor location,  a very simple 
linear combinations of signals can be established to map the activity in 
each auditory cortex.  In effect we make from each 37 channel sensor array 
a Virtual Sensor (VS) which registers the activity in the adjacent auditory cortex~\cite{Liu2}.

MEG is a completely non-invasive method of measuring the distribution 
and time dependence of the magnetic field outside the skull.  
Just like the more conventional Electroencephalography (EEG) it allows to
time-resolve neuronal activity on the scale of 1ms~\cite{Creu}. Its main 
advantage over  scalp-EEG is that the skull and the scalp are transparent to 
the magnetic field and, therefore, an external measured magnetic field is
not distorted by radial conductivity effects. Furthermore, magnetic fields
outside the skull are generated predominantly by the currents tangential
to the surface of the head. The cortical currents are perpendicular 
to the surface of the cortex but almost $70\%$ of the human cortex is folded
into fissures which makes these currents effectively tangential to the skull
and, thus, accessible to MEG.
The above aspects of MEG make it particularly suitable 
for studying the spatio-temporal characteristics of the brain dynamics 
(e.g.~\cite{Jirs}). The details of the MEG experiments used to generate the 
data analysed in this paper are presented in Section 2.

In MEG the response to a stimulus is represented by the time-series, 
one time-series for each channel.
We use the repetition of identical stimulus 
presentations (commonly refer to as trials or epochs) to compute statistical 
measures of correlations or of complexity. 
We use the mutual information (MI)~\cite{Swin}, a concept related to entropy, 
to characterise the correlation 
between the two time-series representing the left and right auditory cortex
activity in a single trial. An outline of the corresponding   formalism, 
including a very useful generalization of MI, is given in Section 3.      
This formalism is then used in Section 4 to study the long-range 
cortical correlations induced by left-ear, right-ear and binaural auditory
stimulations. The paper ends with some concluding remarks.

\section{Description of experiment}

The measurement of the minute magnetic field generated by the coherent activity 
of many millions of neurons can be recorded almost routinely today,  
using super Quantum Interference Devices (SQUIDs) operating within shielded 
environment~\cite{Hama}.  The most advanced instruments today have well over 
100 SQUIDs allowing for a fairly dense coverage of sensors all round the head.
In this work we will report a study performed with the twin MAGNES system 
of Biomagnetic Technologies inc. (BTi) in San Diego.  This system has two 
separate dewars each with 37 first order gradiometers. 
During the experiment,  the subject's head was resting on the bottom dewar, 
while the top dewar was placed over the opposite temporal area.  Five healthy 
male volunteers (age: 37.8$\pm$9.7) gave their informed consent to participate 
in two experiments.  Four subjects (JD, JL, FB and RB)
were right handed, two of them (FB and RB) were twins
and one subject (DB) was left handed.
The first experiment, was in two parts (Ex1a and Ex1b),
with a second experiment, Ex2, performed between Ex1a and Ex1b. 
The second experiment used similar auditory tones in a standard GO/NOGO 
symmetric avoidance protocole.  
For the purpose of this study the details of Ex2 are not relevant,
other than it was long and it involved auditory stimuli which determined 
whether or not a movement was to be made or withheld.  
For more details see~\cite{Liu2}.   The subject maintained 
the same position throughout Ex1 and Ex2, which was fixed as follows:  A 
standard auditory evoked response was first obtained from stimuli delivered to
both ears. This response is termed M100,  
it is the magnetic analogue of the N100,  a peak in the
EEG signal corresponding to the crest of a negative potential~\cite{Creu}.
The BTi software was used to compute and display the average signal 
across 120 single trials,  while the subject remained in place.  The inspection 
of the average signal was used to guide repositioning of the dewars so that the 
prominent M100 peak was captured with the positive and negative fields evenly 
covered by the sensors in each probe.  The procedure was repeated until 
each dewar was well positioned,  usually in one to three placements. 
Two further runs were obtained with this optimal dewar position with 
exactly the same protocol,  but with the stimulus delivered first to the 
left and then to the right ear.  The first part of Ex1 (Ex1a) consisted of 
three runs:  the last dewar placement run with binaural stimuli and the 
two monaural stimulations.  The subject then underwent the more demanding 
and long Ex2.   Immediately after Ex2,  with the subject still holding 
the same position,  experiment Ex1 was repeated 
(for most subjects the binaural tone presentation was omitted). 
 For both the positioning runs and 
the 5 or 6 actual runs of Ex1, the stimuli were 50 msec, 1 kHz tone bursts at 
50dB (10 msec rise/fall and 30 msec plateau). The inter-stimulus interval 
was 1 second  ($\pm$ 20 msec). 
The MEG signal was recorded in continuous mode, 
sampled at 1042 Hz and filtered in real time with 0.1 Hz high pass. 
The analysis to be reported in this paper used two more signals obtained by 
further band-pass filtering in the 1-200 Hz (with notch filters at 50 Hz, 
100 Hz and 150 Hz),  and 3-20 Hz.  

The biomagnetic inverse problem has no unique solution.  This seemingly 
unsurmountable obstacle becomes less formidable when physiological constraints 
are introduced and provided the resolution demanded from the data is limited 
to what is achievable given the sparseness of the sensors and the noise in the 
data.  The extraction of reliable estimates is considerably easier for 
superficial generators,  directly below a sensor array, i.e. the auditory 
cortex in our case.  The requirement to analyze single trials poses new 
problems.  We have 
arrived at a simple but very efficient solution beginning with powerful, but  
computationally demanding methods,  and an analytic transformation of the 
signal, the  ``$V_3$"~\cite{Liu1}.  Comprehensive tests with
computer generated data have shown that a virtual sensor (VS), can be designed 
to respond preferentially to activations of superficial focal source.  
This is similar to earlier work using a template approach~\cite{Liu1},  
but here it has been specifically developed in the context of the 37 channel 
MAGNES system to obtain regions of interest rapidly. For the 
purposes of our investigation a good $VS$ for auditory cortex 
activation can be easily obtained from each probe, provided 
the 37 channels on each side capture symmetrically the dipolar field 
distribution at the peak of the average signal.  
For each probe, we have identified the two channels ($k_1$ and $k_2$), 
which produced the maximum difference at the 
time of the M100 peak, and used them to define the composite VS, 
\begin{equation}
VS^{M100} (t) =  \sum_{j=1}^{37} \, 
\left[ e^{-\left(\frac{|{\bf r}_j-{\bf r}_{k_1}|}{\lambda}\right)^2 } -  
e^{-\left(\frac{|{\bf r}_j-{\bf r}_{k_2}|}{\lambda}\right)^2 } \right]  S_j (t) 
\end{equation}

where $\lambda$ is the characteristic length (we have used $\lambda$ = 0.02 m 
which is roughly the inter-channel separation);  
the results do not depend critically on this value.    
$S_j (t)$ is the MEG signal at time $t$ recorded by the $j^{th}$ channel, 
whose position vector is ${\bf r}_j$.

The coefficients of the expansion are computed at the time of the M100 peak 
in the average signal;  these coefficients are used unchanged for the analysis 
of all single trials.  
The computation of $VS$ is very fast 
and hence $VS$ can be used to scan through all MEG averaged or single trial 
signals very quickly.  For the purpose of this present study the VS output 
from each probe provides a good estimate of the activity in each auditory 
cortex. 
We can therefore use the pair of time-series in each single trial
to study the relationship between the left and right cortex activity.

Fig.~1 summarises the setup (a), shows a typical set of
MEG signals (b) and highlights the area of strong sensitivity for the 
VS corresponding to this signal.

\section{Mutual information and its generalized version}

In an experiment as described above the message about the subsystem $s$
(brain area in this case) behavior is transmitted across the channel of
instruments and procedures, and as a result, is represented by the time-series 
$x_{s}(t_n)$. The subscript $n$ indicates that experiment determines $x_{s}$
at the discrete time points and thus induces a partition of the phase space of
$s$. This time-series maps out the probability $p(j)$ that $x_s(t_n)$ assumes
value characteristic for the $j$th element of the partition.
The average amount of information gained from such a measurement can be 
quantified in terms of the entropy
\begin{equation}
H(X_s)= - \sum_j p(j) \ln p(j),
\label{eq:ent}
\end{equation}
where $X_s$ denotes the whole set of possible messages and the associated
probabilities $(\sum_j p(j) =1)$ for the subsystem $s$.

If two subsystems, $s1$ and $s2$ are measured simultaneously, as is
the case here, then the corresponding probability distributions are
$p(j_1)$ and $p(j_2)$, and the most relevant one, the joint distribution
$p(j_1,j_2)$. For the combined system, composed of $s1$ and $s2$, the joint
entropy $H(X_{s1},X_{s2})$ has the form analogous to eq.~\ref{eq:ent}.
It is easy to verify that 
\begin{equation}
H(X_{s1},X_{s2}) \le H(X_{s1}) + H(X_{s2})
\label{eq:jent}   
\end{equation}
and the equality holds only if $s1$ and $s2$ are statistically independent,
i.e., $p(j_1,j_2)=p(j_1)p(j_2)$.
The quantity 
\begin{equation}
I(X_{s1},X_{s2})=H(X_{s1}) + H(X_{s2}) - H(X_{s1},X_{s2})
\label{eq:mi}
\end{equation}
thus evaluates the amount of information about one of the subsystems resulting
from a measurement of the other and is therefore called the mutual information.
Generalization of this concept to a larger number of subsystems is
straightforward and is known as redundancy~\cite{Fras}.  

A question of fundamental interest, especially in the context described 
in the Introduction, is whether the spatiotemporal correlations between
the subsystems are caused by spatial uniformity or by information
transport. Information transport may lead to time-delayed effects in the
synchronization of correlations. Such effects can easily be quantified by
calculating the time-delayed mutual information between measurements of the 
two subsystems at different times. The corresponding prescription retains of
course the structure of Eq.~\ref{eq:mi}; only the time-series $x_{s1}(t)$
needs to be correlated with $x_{s2}(t+\tau)$. The mutual information
$I(X_{s1},X_{s2};\tau)$ then becomes a function of the time-delay $\tau$.     
It may display maximum at a certain finite value of $\tau$. This value of 
$\tau$ thus provides an estimate on the time needed for the information to be
transported from the subsystem $s1$ to $s2$.  

There exists~\cite{Reny} an interesting generalization of the concept of
the information entropy. It reads: 
\begin{equation}
H_q(X_s)= {1 \over {1-q}} \ln \sum_j p^q(j)
\label{eq:gent}
\end{equation}

For $q \to 1$ this equation yields the standard information entropy 
[Eq.~\ref{eq:ent}].  The most useful property of $H_q(X_s)$ is that with
increasing $q$ a higher weight is given to the largest components in the
set $\{p(j)\}$. This proves very instructive in studying various aspects of the
phase-space exploration in dynamical systems~\cite{Droz}.
Since normally the largest components are likely to dominate the process of
correlating the two subsystems it seems worthwhile to introduce analogous
generalization at the level of mutual information. In fact, recent
literature~\cite{Pomp} considers such a generalization but mostly for $q=2$ 
and on a formal level, without fully documenting its utility in practical terms.      

By making use of the defining equation~\ref{eq:mi}, $H_q(X_s)$ of
Eq.~\ref{eq:gent}, the corresponding generalized joint entropy   
and allowing the time-delay $\tau$ between the time-series, after simple
algebra one obtains the following expression for the generalized mutual
information:  
\begin{equation}
I_q (X_{s1},X_{s2};\tau) = {1 \over {1-q}} \ln {{\sum_{j_1} p^q(j_1) 
\sum_{j_2} p^q(j_2)} \over {\sum_{j_1j_2} p^q(j_1,j_2;\tau)}}.
\label{eq:qmi}
\end{equation}
This equation constitutes a basis for numerical applications and its utility
will be illustrated in the next Section. 

A reliable estimate of the entropy requires appropriately
accurate sampling rate in order to realistically determine the probability
distribution $p(j)$. For this one needs either a sufficiently long single
time-series representing a phenomenon of interest or, as in the present case
of the relatively short time-series, one needs a sufficiently large ensemble
of such series. When estimating $I_q$ in the latter case one thus faces the
two possibilities: (i) $I_q$ is calculated independently from each time-series
and then averaged over an ensemble or (ii) the ensemble averaged probability
distribution is used in Eq.~\ref{eq:qmi}. Obviously, in general the two
operations are not equivalent for this simple reason that the logarithm and 
the sum do not commute. It is quite natural to expect that the prescription
(ii) is more appropriate as it results in a smoother behavior
already on the level of probability distributions, and thus the final result is
to a lesser degree contaminated by artificial noisy fluctuations.
This statement can be confirmed by explicit numerical verification.

From a general point of view one note of caution is also needed
at this point regarding $I_q$. For $q > 1$ it may happen that it assumes
small negative values and the fact that it reaches a zero value does not
automatically mean that the subsystems are statistically independent. 
An inverse implication still holds, however, as for $q=1$: Subsystems which 
are statistically independent lead to $I_q = 0$. Also, the
positive value of $I_q$ for any $q$ means that the subsystems are not
independent. What in this connection is important for us is that the above
peculiarity of $I_q$ for $q > 1$ may apply to the region of very weak
correlations only.

\section{Results}

Little can be extracted from a single pair of time-series.  We need to
consider the ensemble of single trials.  We first establish the notation.

From here on we will restrict our attention to the VS output computed as 
described in Section 2.  Each run is represented by two sets of the time 
series covering the 1s long time-interval $x^{\alpha}_L(t_n)$ and
$x^{\alpha}_R(t_n)$  ($n=1,...,1042$,  corresponding
to the left (L) and the right (R) hemisphere, respectively.   
The sampling rate is 1042 Hz,  so $t_{n+1}-t_n=0.96$ms).
The superscript $\alpha=1,...,120$ labels the single trials in each experiment.
The time-series are consistently centered such that the onset of the stimulus 
corresponds to $n=230$.
Fig.~2 shows three typical, randomly selected, single-trial raw 
time-series together with the average 
\begin{equation}
x_{L,R}(t_n)= {1 \over N} \sum_{\alpha} x_{L,R}^{\alpha}(t_n)
\label{eq:ave}
\end{equation}
over all $N=120$ trials for the left (a) and right (b) hemisphere signals, for 
one subject (JD).   
     
It is difficult to identify the stimulus onset from the raw single-trial
signal,  although a relationship between the peak of the average response 
can be seen in some of the single trials.  For a more detailed discussion 
about the relationship between the average signal and the average 
see~\cite{Liu1,Liu2}. It is clear that the single trial activity
is not dominated by the stimulus. 
Since the background brain activity is not time-locked to the stimulus 
it is averaged out after summing
up a sufficiently large number of identical trials.    
The average over the full set of our 120 consecutive trials exhibits a 
pronounced M100 peak centered at around 100ms after the stimulus onset.
At the time of the M100 a number of generators are active;  
our sensor positioning and the VS analysis in each hemisphere 
disentangles from the MEG data the local 
collective neuronal response at the superficial part of the auditory cortex. 
Interestingly, even though the stimulus is applied 
asymmetrically (left ear) a similar (but not identical) structure is detected
on both hemispheres. This is consistent with the known auditory pathways which 
are less seggregated on the contralateral side than in other sensory modalities,
namely the visual and somatosensory;  an additional contribution may arise 
from long-range interaction between the two cortical auditory areas,  
which are also known to be heavily interconnected via the corpus callossum.       

We first explore the variation of mutual
information between the two hemispheres both, as a function of the time-delay
$\tau$ and as a function of the frequency. The frequency spectrum
of the input data series $x_{L,R}(t_n)$ is determined by their discrete Fourier
transform as
\begin{equation}
X_{L,R}(k) = \sum_{n=1}^N x_{L,R}(t_n) \exp(2 \pi i n k/N),
\label{eq:ft}
\end{equation}     
$X_{L,R}(k)$ being the complex numbers
$(X_{L,R}(k) = \vert X_{L,R}(k) \vert \exp(i\eta (k))$.
By inverting this transformation in a reduced interval 
$\langle K- \Delta K/2,K+ \Delta K/2 \rangle$ of discrete frequencies $k$
one obtains the filtered series $x_{L,R}^{K,\Delta K}(t_n)$
spanning the frequency window $\Delta K$ centered at $K$: 
\begin{equation}
x_{L,R}^{K,\Delta K}(t_n) = {1 \over \Delta K}
\sum_{k=K-\Delta K/2}^{K+\Delta K/2} X_{L,R}(k) \exp(-2 \pi i n k/N).
\label{eq:wind} 
\end{equation}

Determining a minimum value of $\Delta K$ which can safely be used in the
present context requires some care. The point is that its too small value
may generate artificial correlations in the mutual information of filtered
time-series. In an extreme limit, one frequency components will always be 
correlated in some way. What preserves or washes out 
the correlations in finite frequency windows is the relation among 
amplitudes of different frequency components. We
determine a reasonable minimum value of $\Delta K$, such that no artificial
correlations are induced, self-consistently:  we make use of the surrogate time-series of the
original ones. The surrogates are obtained by randomizing the phases
$\eta(k)$ of $X_{L,R}(k)$ and making use of Eq.~\ref{eq:wind}.
This operation preserves the power spectrum of the original series. 
By calculating the mutual information of the so generated surrogates 
of $x_L(t_n)$ and $x_R(t_n)$ we find $\Delta K$ equivalent to 4 Hz as an
appropriate minimum frequency window for our data. Below this value some
correlations may show up even on the level of surrogates.

Fig.~3 shows the landscape of the mutual information $(q=1)$ in the time-delay
$\tau$ and in the frequency window of 5 Hz centered at the value indicated.
This is one of the experiments on subject JD. The results of the other 
experiments for the same subject look similar.

When making use of Eq.~\ref{eq:qmi}, here, as well as in the whole following 
discussion, a grid of 10 bins covering an interval of variation of both,
$x^{\alpha}_R(t_n)$ and $x^{\alpha}_L(t_n+\tau)$ is introduced. This guarantees
stability of the results. For a given experiment the three different 
probability distributions entering Eq.~\ref{eq:qmi} are evaluated by 
superimposing histograms corresponding to all the time-series 
$(\alpha=1,...,120)$ and then the logarithm is taken.
As mentioned before, one could also calculate MI for each $\alpha$ separately
and then average over $\alpha$ but for the present data
such a procedure turns out highly unsatisfactory in terms of statistics;
it results in a much higher level of noisy background fluctuations.

The MI displayed in Fig.~3 is calculated from the whole 1s $(n=1,...,1042)$
time-interval. Its specific $\tau$ dependence will be discussed in full 
detail later and Fig.~3 is basically supposed to illustrate the frequency 
localization of significant correlations. As it is clearly seen, such
correlations are mediated by the low-frequency (up to 20 Hz) activity.        
This picture turns out to be subject independent. The amplitude of MI is
found to depend from subject to subject, however. For certain subjects the
correlations are so week that they are hardly identifiable on the level
of $q=1$ MI. For this reason we first explore a possible advantage of using the
generalized MI as allowed by Eq.~\ref{eq:qmi}. According to the 
above frequency localization, and in order to make the following study more 
transparent, all the time-series used will be filtered to the frequency window
between 3 and 20Hz. Furthermore, since correlations are mainly connected with 
appearance of the M100 peak, the time-series will be truncated to the
interval between $i=230$ and $i=491$. This covers 250ms starting exactly
at the initial moment of the stimulus.

The benefit of using the higher $q$-MI is documented in Fig.~4. This figure
illustrates the $q$-dependence $(q=1,2,4,6)$ of the generalized mutual 
information for the two examples: of strong correlations (JD) and of weak 
correlations (FB). Clearly, the higher $q$-values offer a much more precise   
estimate of the time-delay $\tau$ at maximum. This originates from the
fact that increasing $q$ gives a higher weight to larger components in 
the probability distribution and this turns out especially important for 
the cases of weak correlations. For this reason a summary of the results
of all experiments, for all five subjects, as displayed in Fig.~5, is done
for $q=6$. A convention used in the corresponding calculation when defining
the sign of the time-delay $\tau$ between $x_L(t_n)$ and $x_R(t_n+\tau)$ 
is such that its negative value means that a relevant excitation in the right 
hemisphere is time-advanced relative to the left hemisphere. Of course, the
opposite applies for positive sign.                      
     
Several conclusions are to be drawn from Fig.~5a.
First of all, the correlations under study are spatially nonuniform
and the information transport between the hemispheres takes about 10ms.
The relative location of the peaks in MI indicates that, at least
statistically,  the contralateral hemisphere drives the response
for all the subjects and conditions studied. This, however, in general 
can only be identified by a parallel analysis of the left versus right ear 
stimulation (binaural is also helpful) of the same subject. 
The point is that for some subjects there are certain asymmetry effects. 
For instance, in JL the ipsilateral hemisphere somewhat overtakes ($\sim$ 5ms)
when the right ear is stimulated but then the contralateral hemisphere 
overtakes even more when the tone is delivered to the other ear, 
so that the relative location of the peaks in MI, corresponding to the 
left and right ear stimulation, respectively, is still preserved. This
asymmetry in JL disappears in the experiment Ex1b, however.
A trace of asymmetry, but in opposite direction, is also visible in JD; again
more in Ex1a than in Ex1b. A likely explanation of those asymmetry effects
is that we are facing a superposition of the two phenomena. 
One is a leading role of the contralateral hemisphere when the tone is
delivered to one ear (either left or right) and the other may originate from
certain subject specific asymmetry in properties of the left and right auditory
areas. The latter kind of asymmetry is known to occur quite
frequently~\cite{Creu}.

Figs. 5b and 5c illustrate the same quantities for the time-intervals just
before the stimulus onset (-230 -- 0 ms) and soon after the M100 period
(251 -- 500 ms), respectively. The picture changes significantly.
Except for JD the correlations essentially disappear. JD seems to display
certain permanent inter-hemispheric correlations
but here they are considerably weaker and always driven by the
left hemisphere. This supports the claim that the correlations under study
are primarily associated with the stimulus.

Another interesting quantity is the strength of information transfer 
between the hemispheres. 
This characteristics measured in terms of the MI-excess over background 
is largely invariant for a given subject (similar for different experiments). 
It is, however, strongly subject dependent and ranges between very
pronounced (JD) and rather weak (FB, RB). A related question that emerges
in this connection is whether this effect results from different strength
of the coupling between the hemispheres or, whether this is due to the fact
the local M100 excitations differ in their degree of collectivity.
That the second possibility is more likely to apply here can be concluded from
Fig.~6 which shows the averaged (over 120 trials) MEG time-series for JD, DB,
JL and FB (RB looks similar to FB). This figure illustrates both, the left and
the right hemisphere responses generated by the left, right and binaural 
stimulations. The results shown also display left/right hemispheric asymmetry
for JD and JL, oriented consistently with the results of Fig.~5a.
The magnitude of amplitude of the so quantified response reflects a degree of
neuronal synchrony developing the M100 complex in each case and this amplitude
goes in parallel with the strength of the information transport (Fig.~5a).
This, in a sense, is natural since the amount of information to be communicated
results from the original local collectivity.
It is also consistent with the low-frequency origin of
inter-hemispheric correlations as illustrated in Fig.~3. Localization in
frequency means higher synchrony and more determinism, 
and these, in general, constitute preferential
conditions for the long-range inter-hemispheric correlations to occur.  

Such conclusion gets further support from the structure of the power spectrum
(squared modulus of the Fourier transform) of the time-series.
These power spectra are calculated from the original time-series $x_{L,R}(t)$
representing the whole specific experiment lasting 120s and are 
shown in Fig.~7 for the two extreme cases, JD and FB, respectively.
The cases of stronger correlations (JD) are accompanied by the power spectra
that have the lower-frequency part significantly
amplified relative to the cases of weak correlations (FB, RB). 
The opposite applies to the high-frequency region. The weak correlations are
thus connected with a more noisy dynamics which acts destructively on local
coherence and, consequently, on long-range correlations.
It is however interesting to notice that even in this case
the power spectrum is not completely flat as for the white noise phenomena
but shows a nice "$1/f$"-type (straight line of finite negative slope
in the log-log scale) behavior~\cite{Dutt}. For JD this behavior is not
that nice but the deviation seems to be largely attributable to the permanent
activity at 8 Hz ($\alpha$-rhythm), consistent with the previous discussion.
In fact, this kind of power-spectrum one may anticipate
already by looking at the M100 waveform seen in the average
signal (Fig.~6) and taking into account its functional similarity to the QRS
complex of the electrocardiogram. This complex develops the inverse power-law
spectrum which some interpret in terms of the fractal character of the cardiac
His-Purkinje conduction system~\cite{West}.
In the present context this perhaps signals that the evolution of the M100 
may be governed by
a phenomenon of self-organized criticality~\cite{Bak} which is a more
catastrophic form of collectivity and is generated by a fractal 
(scale-invariant) 'avalanche'-like process.
Interestingly, a new class of neural networks based on {\sl adaptive
performance networks}~\cite{Alst} shows exactly this type of power spectra.
It also allows some local deviations from this behavior and those deviations
result from certain subject specific stronger activity at some frequency.
This model involving the elements of self-organized criticality can be
trained~\cite{Stas}  to react 'intelligently' to external sensory signals.

Of course, because of a strong permanent brain's activity, it is much more
difficult to precisely disentangle from the background a contribution 
to the MEG power spectrum of a specific structure such as M100.
Therefore, the discussion related to Fig.~7 can only be  treated as an 
indication that the M100 complex may itself obey an inverse power-law.
On the level of average time-series (Fig.~6) it can also be verified that
it does although the statistics is then poorer. The related questions are, 
however, not the central issue of the present paper and will be the subject
of our independent, more systematic, future study, both on the experimental
as well as on the theoretical level.

Finally, on a way towards understanding the mechanism of inter-hemispheric 
correlations, it is instructive to look at MI between $x^{\alpha}_L(t)$
and $x^{\alpha + \Delta}_R(t)$ for $\Delta \ne 0$.
Fig.~8 shows that, surprisingly, such correlations are much weaker for both
subjects (as well as for all remaining).
This result indicates that what actually correlates the opposite hemispheres
in the present context is not just an independent appearance of M100 in both
hemispheres but the real inter-hemispheric information transport which 
projects one M100 into another and thus induces certain similarity between
them. They are thus functionally related and this is what the mutual information
reflects. On the other hand the specific evolution of M100 with respect to
consecutive trials must involve nondeterministic  elements which make the
above, translated correlations much weaker.
This means that only the global aspects of M100 are time-locked to the stimulus;
a detailed  'microscopic' evolution turns out largely stochastic. 
In fact, such a picture is again consistent with the phenomenon of 
self-organized criticality.
    
Besides the inter-hemispheric information transport discussed above
there potentially exists another mechanism capable of introducing a time delay
in the mutual information,  namely a common driver which independently activates
each path at separate times. Such a mechanism does not, however, seem to be able
to explain such a significant change of correlations as shown in Fig.~8.

\section{Conclusions}

The present study provides a clear quantitative evidence for two levels
of dynamical cooperation in the brain auditory processing.
One is the local hemispheric collective response, reaching its maximum at
about 100ms (M100) after a stimulus onset. 
An interesting emerging aspect of this excitation is that its 
only global characteristics are time-locked to a stimulus.
The underlying neuronal degrees of freedom involved are likely 
to significantly differ from trial to trial and a possible scenario
potentially able to reconcile these two aspects is self-organized criticality.
In fact, the structure of the corresponding power spectra is consistent
with such a scenario. It also goes in parallel with our recent suggestion
that the single trial activity induced in the auditory cortex by a simple
tone cannot be treated as a deterministic response emerging from a noisy
background~\cite{Liu1,Liu2}.
The second level of cooperation is the communication between the two
hemispheres. The most conclusive in this connection are the monaural
stimulations. The analysis then shows that, at least statistically,
the contralateral hemisphere systematically leads by 10-20ms. The mechanism
of this communication carries the signature of (delayed) synchronization
and thus can be hypothesised as a direct information transport
between the hemispheres.

An independent conclusion to be drawn from our study is that
the mutual information (MI) and, especially, its generalization, provides
a useful and statistically appropriate formalism for studying the temporal 
aspects of correlations in complex dynamical systems, even if, as here,
such systems are represented by relatively short time-series.

%\acknowledgments

This work was supported in part by Polish KBN Grant No. 2 P03B 140 10.

%-----------------------------------------------------------------------

\newpage
\begin{center}
{\bf FIGURE CAPTIONS}
\end{center}
%------------------------------------------------------------------------
{\bf Fig.~1.} Sensor arrangement,  signal and sensitivity profile of 
Virtual Sensor.
(a)  Coronal and sagittal views showing the senso arrangement relative 
to the head and brain. (b)  The average MEG signal for tone presentation to 
the left ear in the channels of the left and right probe.  The channels 
with the strongest positive and negative signal are marked for each probe. 
The difference of weighted sums of channels, with weights decreasing with 
distance away from the highlighted channels define the Virtual Sensor. 
(c)  By combining the sensitivity profile (lead field) of each channel 
according to how the channel is weighted in the VS sum,  we obtain the 
sensitivity profile of the VS,  which is clearly focussed in the auditory 
cortex.\\
%------------------------------------------------------------------------
{\bf Fig.~2.} Three randomly selected raw MEG time-series (dashed, dash-dotted
and dotted lines) versus the average over the whole set of 120 of them
for the subject JD and left ear stimulation. Upper part illustrates the right
hemisphere and lower part the left hemisphere behavior. \\ 
%------------------------------------------------------------------------
{\bf Fig.~3.} Time-delayed MI as a function of the frequency  
(frequency window of 5 Hz) for subject JD, left ear stimulation.\\
%------------------------------------------------------------------------- 
{\bf Fig.~4.} Two examples of the $q$-dependence of generalized MI
(for $q=1,2,4$ and 6). Upper part corresponds to JD (strong correlations) 
and lower part to FB (week correlations).\\
%-------------------------------------------------------------------------
{\bf Fig.~5a.} $q=6$ MI as a function of the time-delay for all five subjects
calculated from the time-interval between 0 (stimulus onset) and 250ms.
Left column corresponds to the experiment Ex1a and right column to
the experiment Ex1b.
Solid line displays the response to the left ear, dashed line to the right ear
and dash-dotted line to the binaural (only Ex1a) stimulation.\\
%--------------------------------------------------------------------------
{\bf Fig.~5b.} The same as Fig.~5a calculated from the 230ms long time-interval
starting 230ms before the stimulus onset. \\
%--------------------------------------------------------------------------
{\bf Fig.~5c.} The same as Fig.~5c calculated from the time-interval
between 251ms and 500ms. \\
%--------------------------------------------------------------------------
{\bf Fig.~6.} The avaraged MEG time-series over all 120 trials for four 
different subjects corresponding to the left ear (LE), right ear (RE) 
and binaural (B) stimulation.
The solid line displays the left hemisphere and the dashed line the right
hemisphere response.\\  
%--------------------------------------------------------------------------
{\bf Fig.~7.} Power spectrum of the full MEG time-series. 
The upper part illustrates a typical behaviour for JD 
and the lower part for FB. 
The deep at 50 Hz is due to the notch filter applied at this frequency. \\

%--------------------------------------------------------------------------
{\bf Fig.~8} Two examples (for JD and FB) of $q=6$ MI between the time-series 
representing different trials, i.e., $x^{\alpha}_L(t)$ is correlated with
$x^{\alpha + \Delta}_R(t)$. The solid line corresponds to $\Delta=0$ 
(original case), $\Delta=1$ to the dotted line, $\Delta=4$ to the dash-dotted
line and $\Delta=10$ to the dashed line. \\ 
%--------------------------------------------------------------------------    

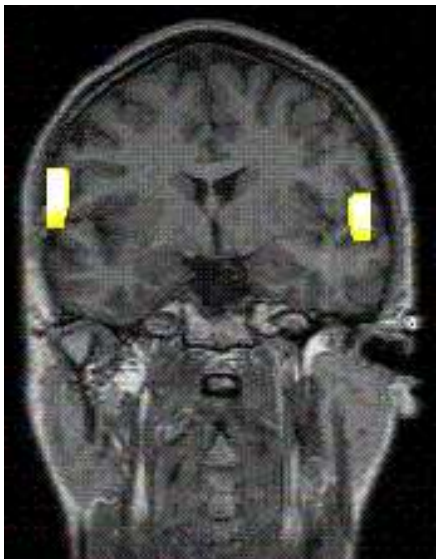
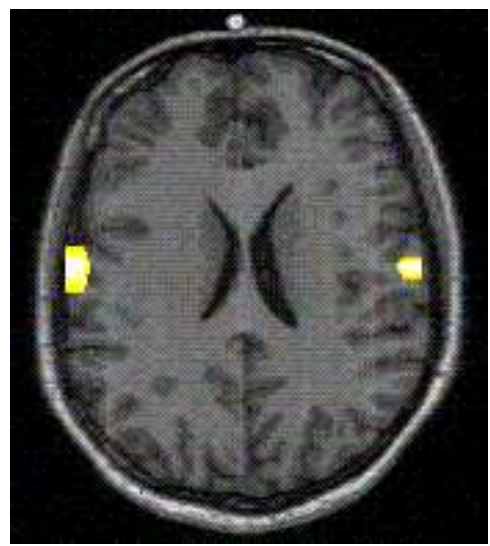
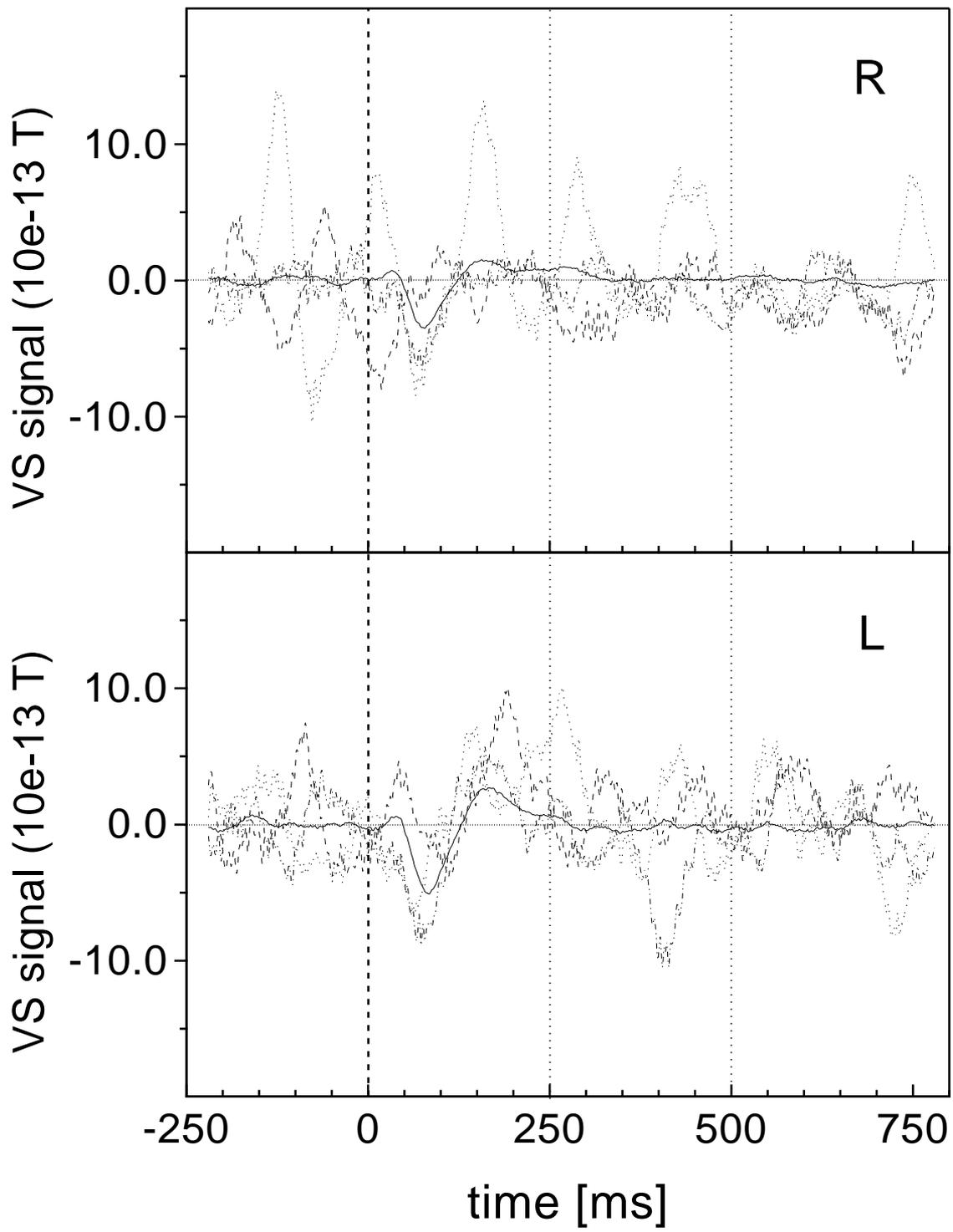
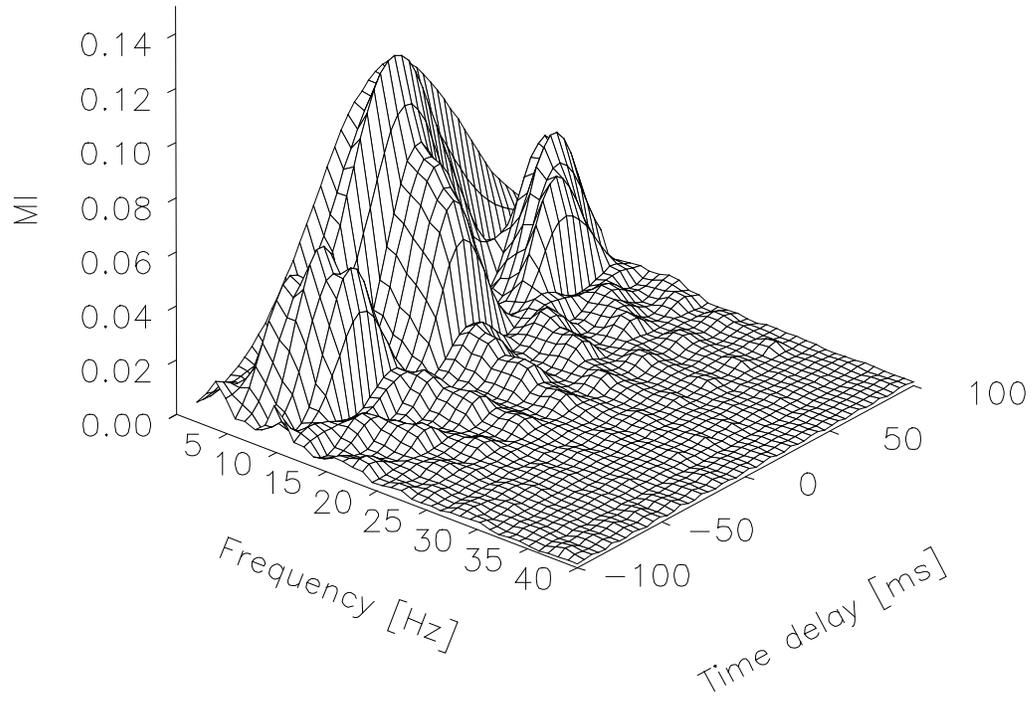
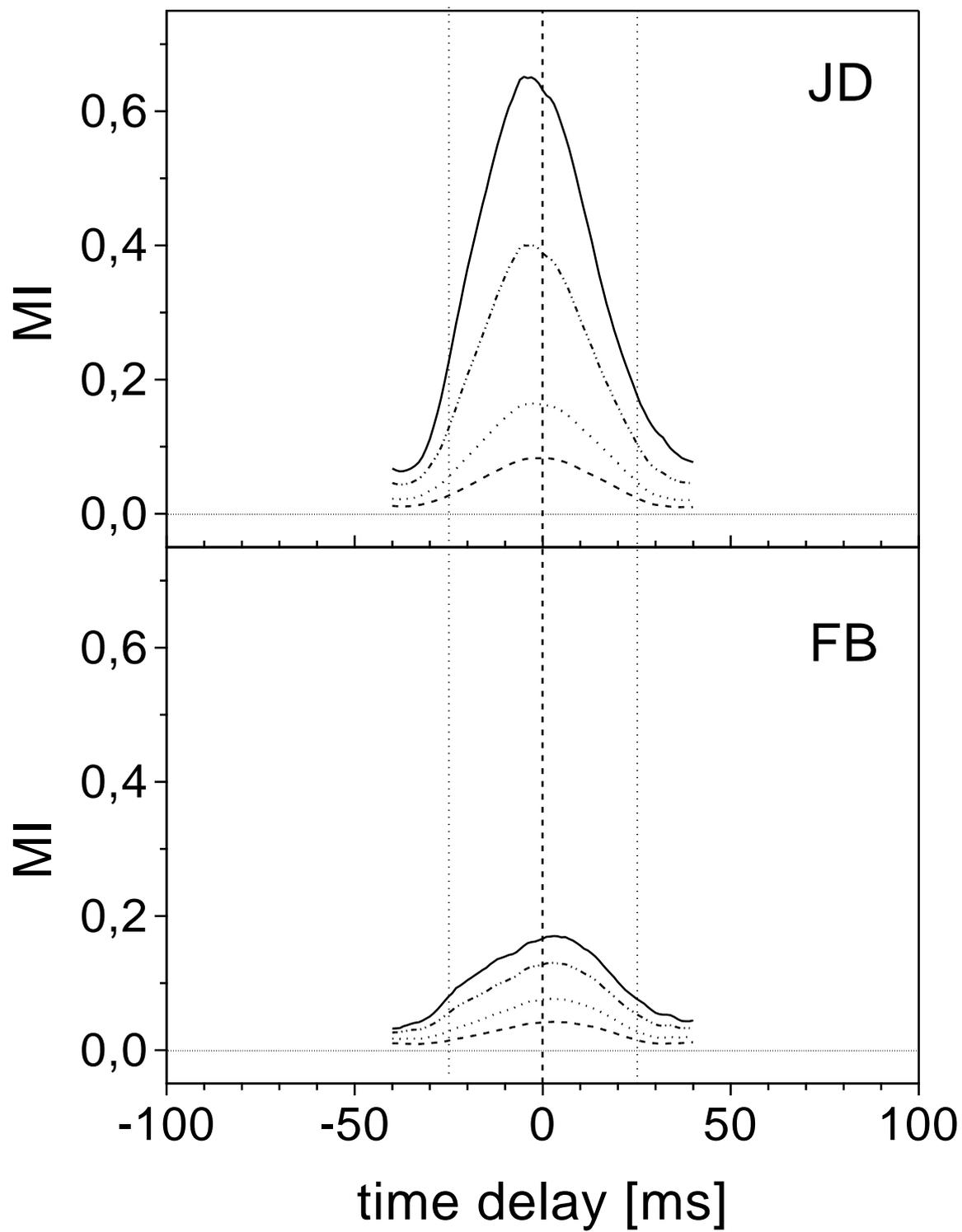
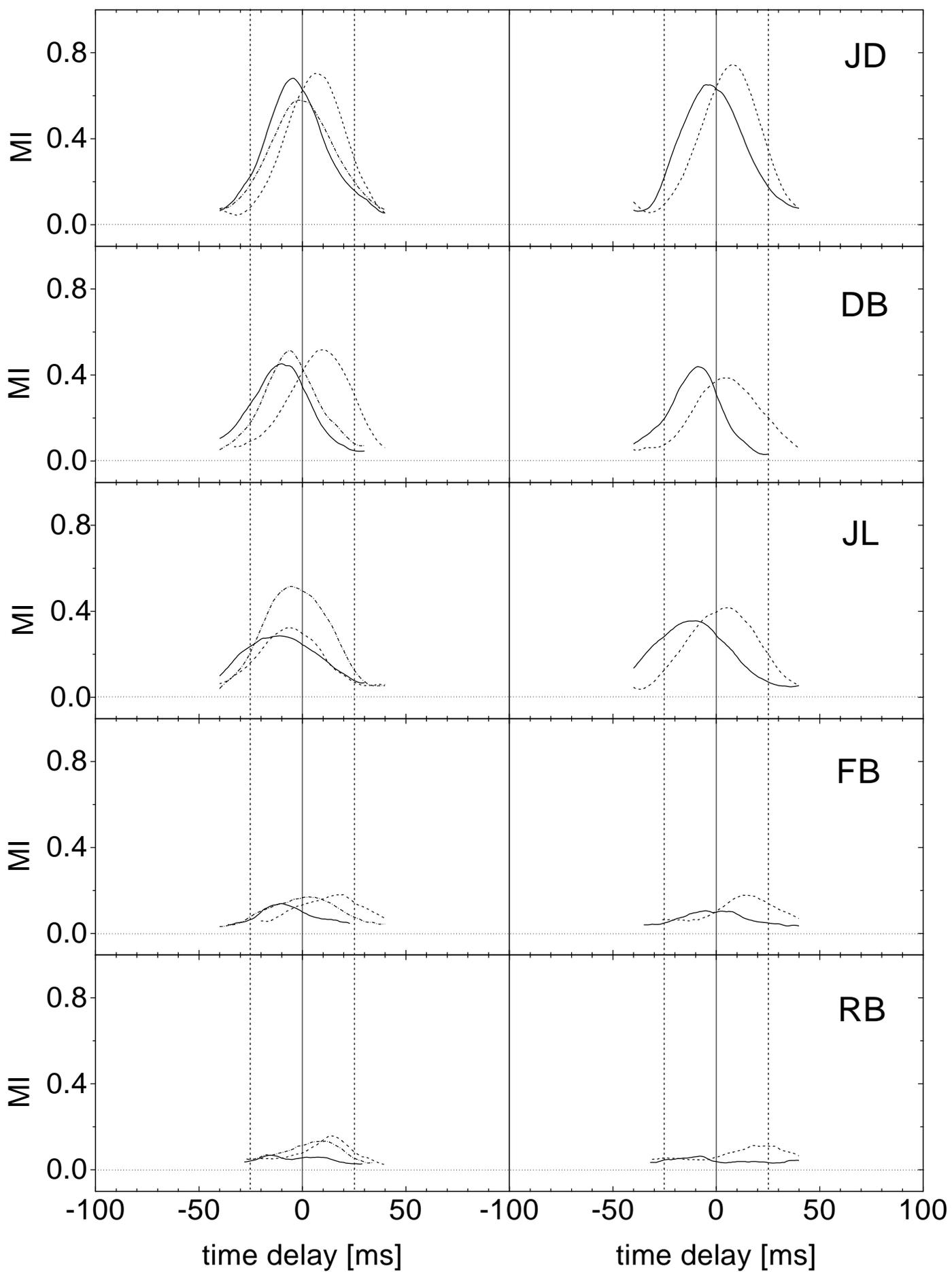
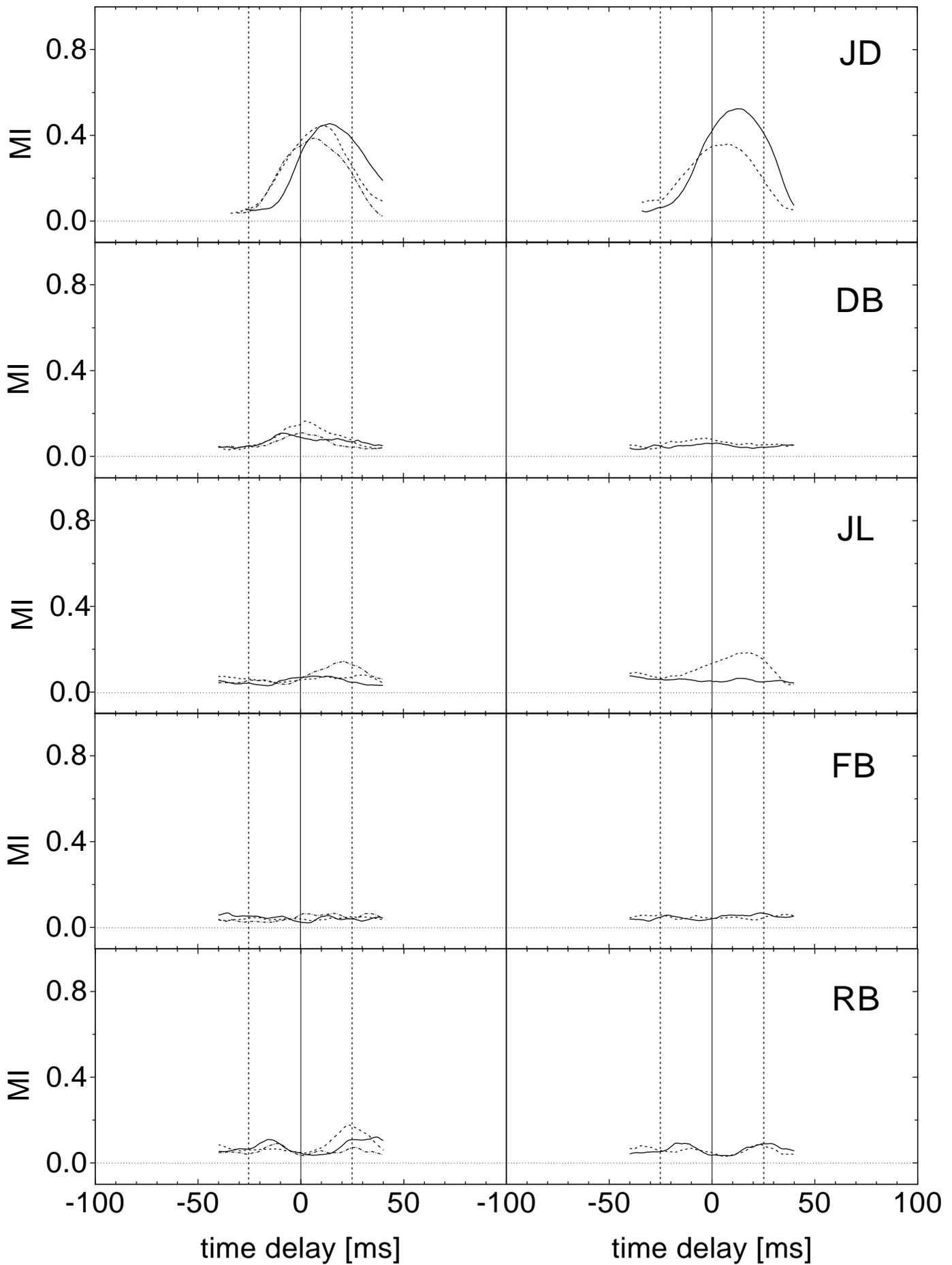
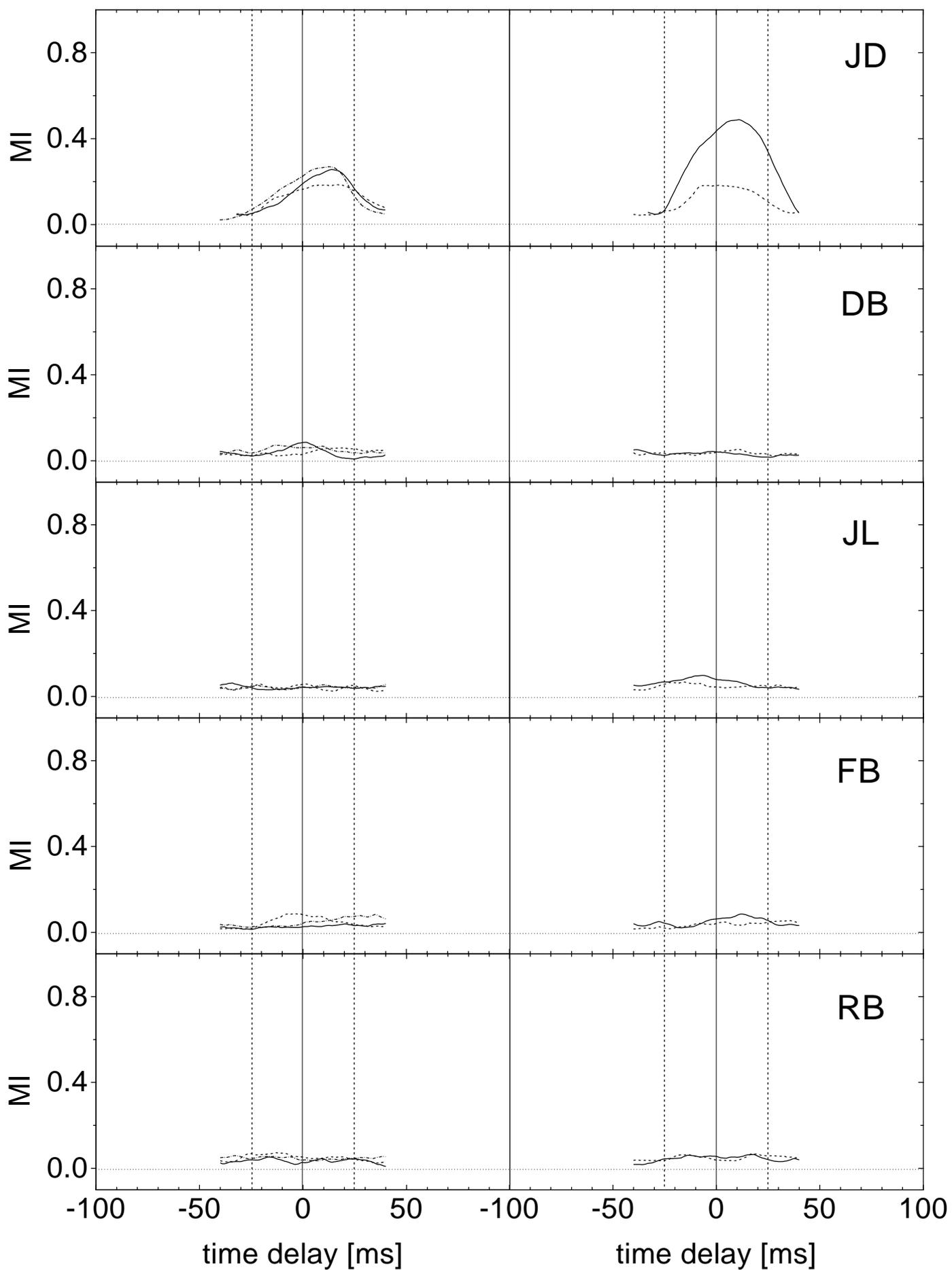
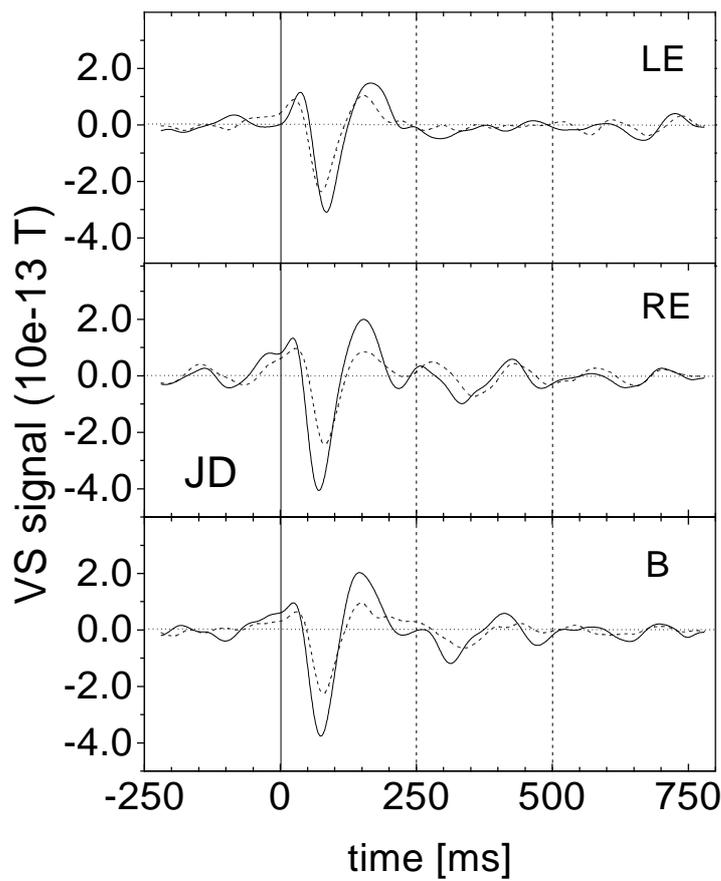
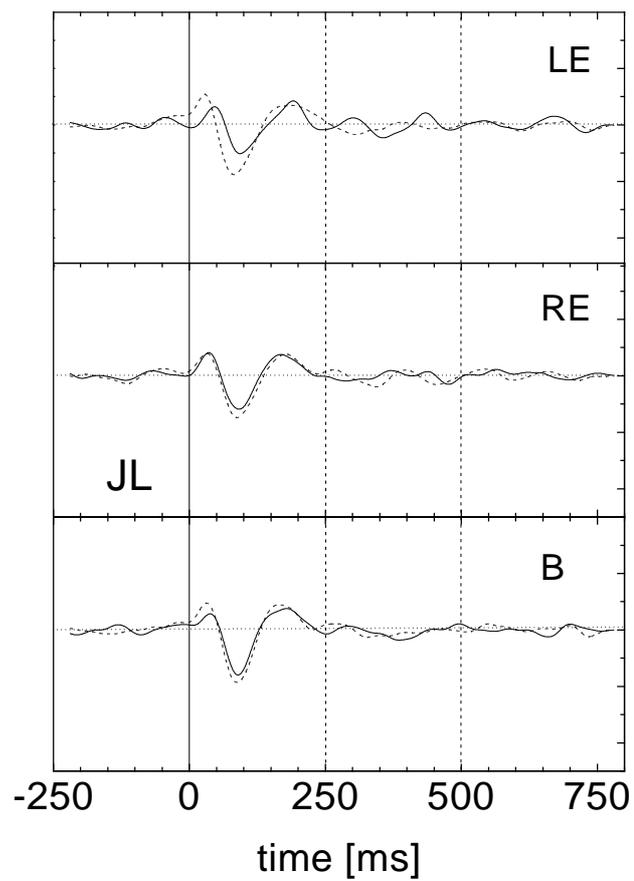
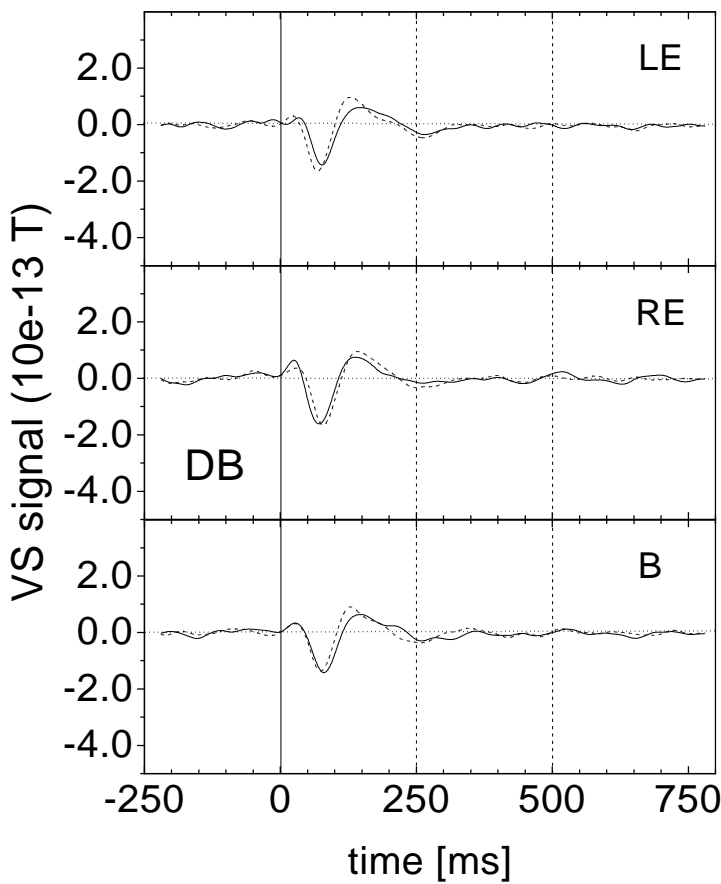
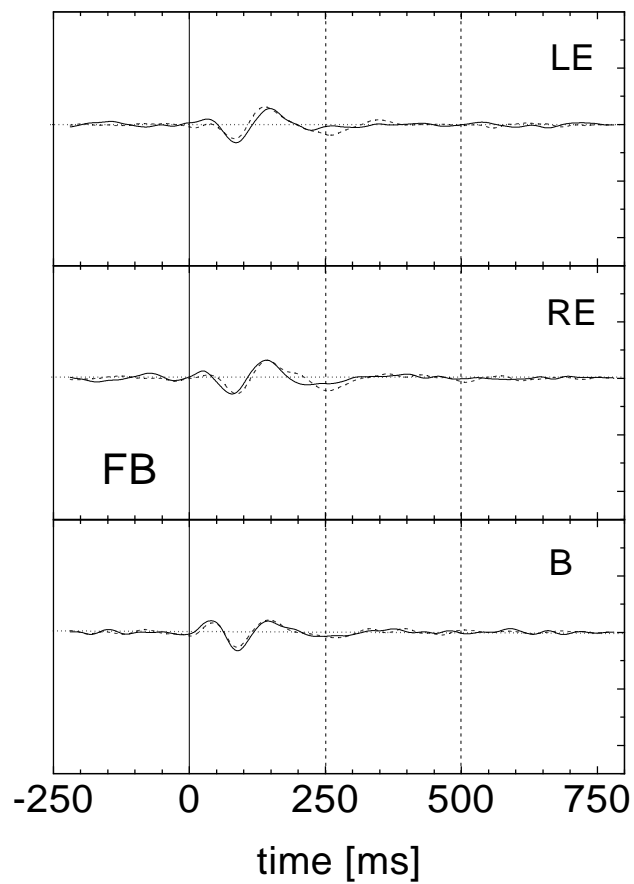
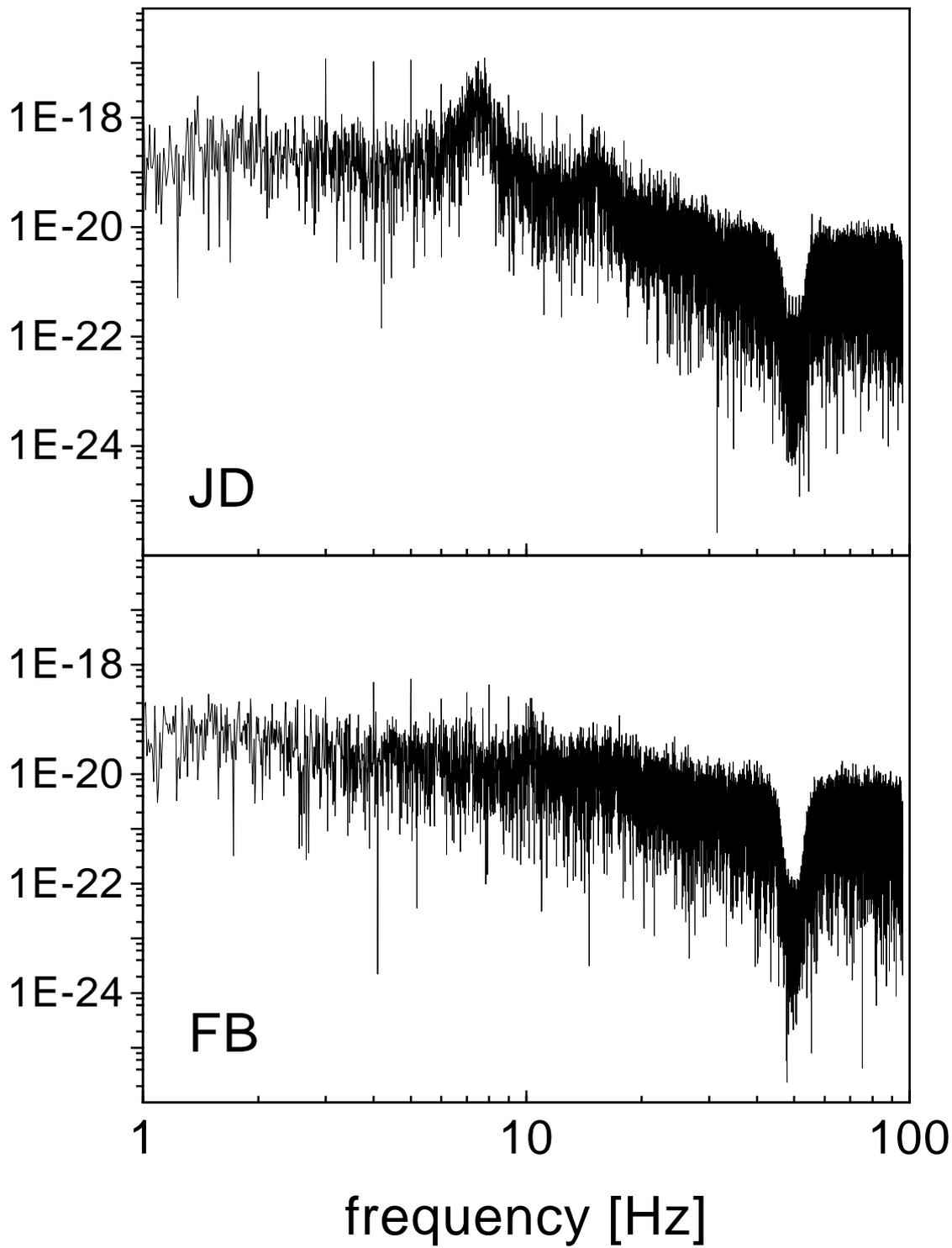
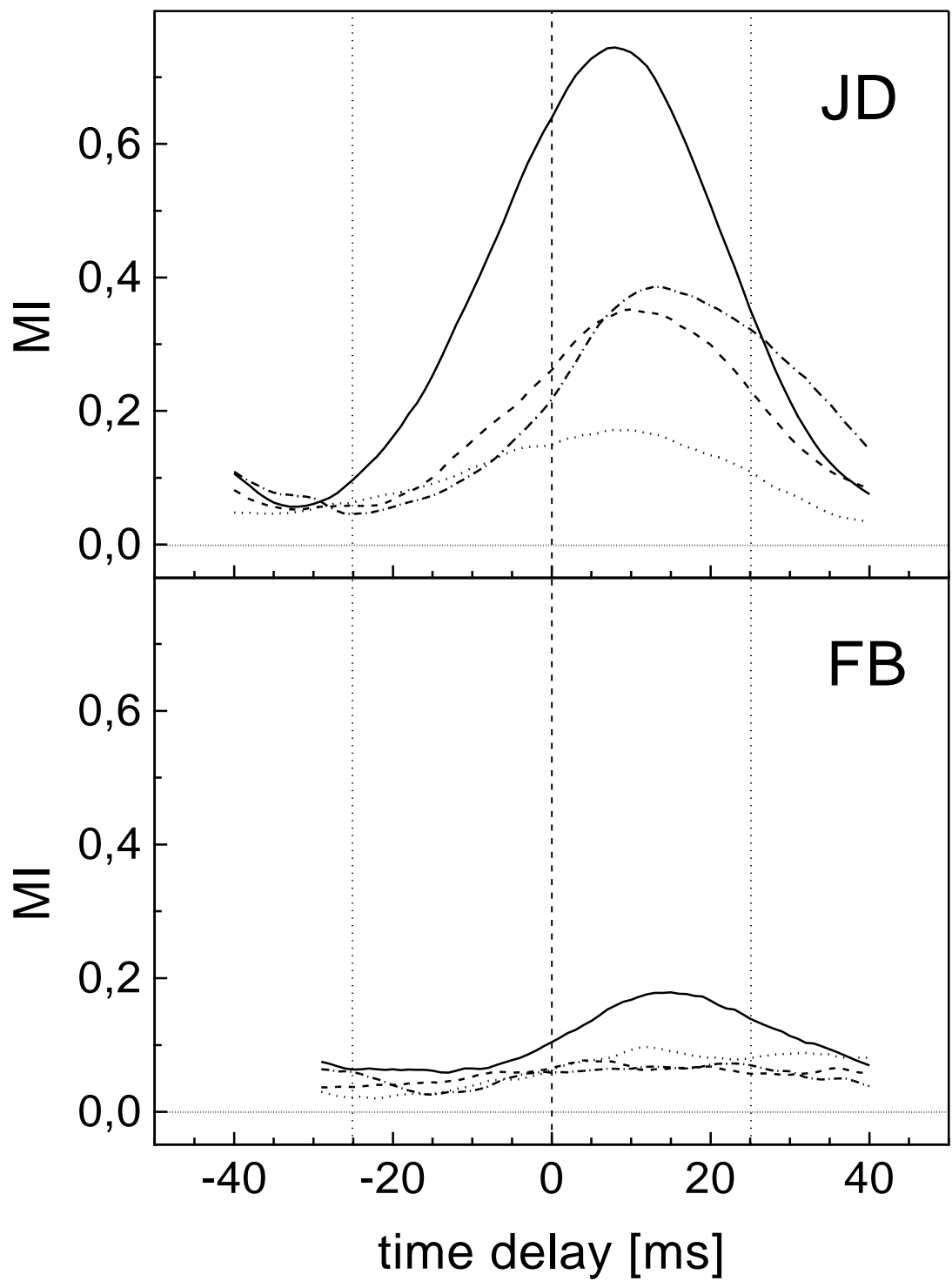
\end{document}